\documentclass[times,sort&compress,3p]{elsarticle}
\journal{Journal of Multivariate Analysis}
\usepackage[labelfont=bf]{caption}

\usepackage{amsmath,amsfonts,amssymb,amsthm,booktabs,color,epsfig,graphicx,hyperref,url}

\usepackage{bbm, subcaption, algorithm, algpseudocode}

\newcommand\bluetext[1]{{#1}}

\theoremstyle{plain}% Theorem-like structures provided by amsthm.sty

\theoremstyle{definition}

\begin{document}

\begin{frontmatter}

\title{On projection methods for functional time series forecasting}

\author[1]{Antonio El\'ias\corref{mycorrespondingauthor}}
\author[2]{Raúl Jiménez}
\author[3]{Han Lin Shang}

\address[1]{OASYS group, Department of Applied Mathematics, Universidad de Málaga, Málaga, Spain}
\address[2]{Department of Statistics, Universidad Carlos III de Madrid, Madrid, Spain}
\address[3]{Department of Actuarial Studies and Business Analytics, Macquarie University, Sydney, Australia}

\cortext[mycorrespondingauthor]{Corresponding author: OASYS group,  Department of Applied Mathematics, Ada Byron Research Building, C/ Arquitecto Francisco Pe\~{n}alosa, 18, 29010, University of Málaga, Málaga, Spain; Email: aelias@uma.es; ORCID: \url{https://orcid.org/0000-0002-3078-549X}; Email address: \url{aelias@uma.es}}

\begin{abstract}
Two nonparametric methods are presented for forecasting functional time series (FTS). The FTS we observe is a curve at a discrete-time point. We address both one-step-ahead forecasting and dynamic updating. Dynamic updating is a forward prediction of the unobserved segment of the most recent curve. Among the two proposed methods, the first one is a straightforward adaptation to FTS of the $k$-nearest neighbors methods for univariate time series forecasting. The second one is based on a selection of curves, termed \emph{the curve envelope}, that aims to be representative in shape and magnitude of the most recent functional observation, either a whole curve or the observed part of a partially observed curve. In a similar fashion to $k$-nearest neighbors and other projection methods successfully used for time series forecasting, we ``project'' the $k$-nearest neighbors and the curves in the envelope for forecasting. In doing so, we keep track of the next period evolution of the curves. The methods are applied to simulated data, daily electricity demand, and NOx emissions and provide competitive results with and often superior to several benchmark predictions. The approach offers a model-free alternative to statistical methods based on FTS modeling to study the cyclic or seasonal behavior of many FTS.
\end{abstract}

\begin{keyword} %alphabetical order
Functional time series \sep
Functional nonparametric \sep
Forecasting \sep
Functional depth \sep
$k$-nearest neighbors \sep 
Electricity demand \sep
NOx emissions
\MSC[2020] 62R10 62G30
 
\end{keyword}

\end{frontmatter}

\section{Introduction}\label{sec:intro}

\bluetext{In the last few decades, technological advancements have simplified and decreased the cost of data collecting and storing processes. This new paradigm has made data scientific to confront not only big data sets but also complex data structures. Functional data analysis (FDA) arises naturally in this context to exploit the information recorded over a continuum such as time or space. The literature has experienced a prominent development of techniques and tools to take advantage of this kind of data. While the monographs of \cite{ramsay2005, ferraty2006, kokoszka2017} provide complete picture of the methodological and theoretical contributions, recent advances in the field can be found in survey papers \cite{CUEVAS2014, GOIA2016, wang2016, ANEIROS2019}.}

This \bluetext{article} addresses the problem of forecasting functional time series (FTS). These are time series $\{y_k, k\in \mathbb{N}\}$ where each $y_k$ is a random function $t\rightarrow y_k(t)$, $t \in [a,b]$. We refer the reader to \citet{hormann2012} for theory and examples of FTS. One-step-ahead forecasting is one of the most important problems in FTS analysis, and it has been addressed by diverse prominent literature \citep[see, e.g.,][]{bosq2000, antoniadis2006,  hyndman2007, aneirosperez2008, hyndman2008, Hyndman2009, aneirosperez2011, Aue2015, Rana2018}. Also, forward prediction of a curve which is already being observed, termed dynamic updating by \citet{hyndmanshang2011}, a topic of current interest \citep{Shang2017, Shang2018}. As many classical methods for time series forecasting, for example, those based on the ARIMA family models, the benchmark methods for FTS forecasting are based on fitting some statistical model of some approximated data generation. The most common is to fit ARIMA or VAR models to the Functional Principal Components Scores \bluetext{(FPC)} of the functions, see FAR model \citep{hyndman2007, Aue2015}. \citet{KLEPSCH2017} considered Functional Moving Average (FMA) and FARMA. \cite{shangARFIMA2020} and \cite{LiShang2020} consider functional autoregressive fractionally integrated moving average (ARFIMA). Here we discuss an alternative model-free approach that addresses both one-step-ahead forecasting and dynamic updating in a unified way.

The literature of artificial intelligence and machine learning for time series forecasting has gained considerable prominence over the last decade \citep[see, e.g.,][]{makridakis2018}. Besides the lack of an underlying model, many of these methods exhibit interesting features compared with classical statistical methods, including their ability to capture nonlinearities \citep{tsfknn}. Among the most powerful machine learning technics used for time series forecasting, we must mention artificial neural networks \citep{ZHANG1998} and $k$-nearest neighbors (KNN). The simplicity of implementing the algorithm and its high accuracy have popularized KNN. It has demonstrated to be a strong contender in forecasting competitions \citep{Martinez2019}. The method can be schematized as follows: 
\begin{enumerate}
	\item Consider  $T$ observations, $x_1, \dots, x_T$, from a time series and a prediction horizon $h$. Our goal is to forecast $p = (x_{T+1},\dots, x_{T+h})$, i.e. the time series $h$ periods ahead. 
	\item Consider all the possible observed  $(E+h)$-tuples $(x_{i-(E-1)}, \dots, x_i, x_{i+1},\dots, x_{i+h})$. This means $E\leq i\leq T-h$. 
	\item Denote ${\bold x}_i = (x_{i-(E-1)}, \dots, x_i)$ and its \emph{projection} $P({\bold x}_i) = (x_{i+1},\dots, x_{i+h})$. Note that $p$ is the projection of ${\bold x}_T$.
	\item Find the $k$-nearest neighbors to ${\bold x}_T$ from $\{{\bold x}_i, E\leq i\leq T-h\}$.
	\item Predict  $p$ by a convex combination of the projections of the $k$-nearest neighbors.
\end{enumerate}
In addition to the distance function used to find the nearest neighbors and the convex combination used for predicting, KNN involves setting the sometimes called \emph{embedding dimension} $E$ and the number of neighbors denoted by $k$. The simplex projection \citep{sugihara1990} and the S-map projection \citep{smap1994} are methods related to KNN that provide skillful forecasts, free of $k$. 

Moreover, these model-free methods have outperformed any forecast based on models \citep{perretti2013}. Surprisingly, \bluetext{functional versions} have not been developed for \bluetext{FTS. In contrast, in the context of FDA, the KNN method has been explored and theoretically supported towards regression problems \citep{florent2009, biau2010, kudras2013, Lian11} or functional classification problems \citep{cerou2006, Hubert2017}. \citet{biau2010} introduces asymptotic results when the response variable has a finite dimension. \citet{kara2017} provides asymptotic results that, exceptionally, allows for an automatic choice data-driven of the number of neighbors for several operators, among them functional regression with a scalar response. \citet{Lian11} proves the consistency of the KNN regression estimate when both dependent and independent variables are functions. Recently, \cite{Ferraty2021} use the nearest
neighbours method for bandwidth selection on a problem of scalar-on-function local linear regression.} 
Remarkably, \cite{Zhang2010} addresses the problem of forecasting final prices of auctions via functional KNN. Roughly, their forecasting is based on the $k$-nearest price curves to the item being predicted and their corresponding final prices. We consider dynamic updating in line with these authors. For example, if we predict afternoon electricity demand, we use afternoon data from days with morning data near to the morning of the day being predicted. For one-step-ahead forecasting, we mimic the projection procedure inherent to KNN. For example, for tomorrow's forecasting electricity demand, we consider next-day demands of daily curves near to today. Moreover, we take ideas of \cite{sugihara1990}, and we do not only consider the nearest daily curves but those that surround the most recent data from above and below, when this is possible, to select a neighbor that is representative of its shape and its magnitude. This is the key to produce coherent band predictions, in which the observed part of the day being predicted is inside the band delimited by the curves that we are using for predicting. \bluetext{Whereas the curve to predict may be entirely outside of the band delimited by its $k$-nearest curves. In this vein,} we combine two concepts, nearness and centrality, commonly termed depth in the functional context.

Functional depth \bluetext{\citep{nietoreyes2016, gijbels2017}} has received a great deal of attention since it was introduced by \citet{Fraiman2001}. It has been used for several applications; including classification \citep{saraLopezJuanRomo2006, Cuevas2007, cuestaalbertos2008, Hubert2017, cuesta2017}, outlier detection \citep{Febrero2008, IevaPaganoni2013, arribasromo2014, Narisetty2015, Nagy2017}, populations comparison \citep{pintadoromo2009, lopezpintado2010, Nicholas2015}, and clustering \citep{Singh2016,Tupper2017}. Functional versions of boxplots and other graphical tools based on different depths have also been proposed for visualizing curves to discover features from a sample that might not be apparent using other methods \citep{Hyndman2010, sungenton2011, Serfling2017, daigenton2018}. However, up to our knowledge, the concept of depth has not been used for forecasting. 

Here we introduce a depth-based projection method for forecasting \bluetext{that constitutes a contribution to the literature of nonparametric FDA, an area popularized by \cite{ferraty2006} and that it is actively growing, see the review by \citet{ling2018}}. 
Also, we consider an adaptation to FTS of the KNN briefly schematized above \bluetext{and we follow ideas from \cite{kara2017} to make it automatic}. We also consider several benchmark methods in the field for comparisons. For that, we use daily functions of electricity demand and nitrogen monoxide (NOx) emissions. Both are FTS that have already been considered by the literature. Also, both exhibit a strong calendar-effect \citep{earo2020}, making them appropriate to illustrate the excellent performance of the method discussed here. As we discuss below, the projection methods have been intended to forecast FTS with seasonal or cyclical patterns. Through a simulation study, we investigate the impact on forecast accuracy from the departure of stationarity. We remark that, although we do not assume any stationarity of first or second order, we do not consider applications to FTS with a significant trend. Specifically, we suppose the curves of the FTS fluctuate in a finite band.

This paper is organized as follows. Section~\ref{sec:FCR} describes the projection methods considered here. Section~\ref{sec:results} provides results of comparing the proposed methods with benchmark methods. Section~\ref{sec:conclusion} concludes with a general discussion. 

\section{Projection methods for forecasting}\label{sec:FCR}

Let $\{y_i, i \in \mathbb{N}\}$ be a FTS \citep{hormann2012}. Here, we assume that each $y_i$ is a univariate random function and, without loss of generality, we suppose that its domain is $[0, 1]$.

\subsection{Basic notions}\label{sec:basicNotions}

We consider two scenarios of prediction: 
\begin{itemize}
	\item One-step-ahead forecasting; when we have observed $y_1,...,y_n$ on $[0,1]$, and we want to predict $\{y_{n+1}(t): t\in [0, 1]\}$.
	\item Dynamic updating; when we have observed $y_1,...,y_{n-1}$ on $[0,1]$ and $y_{n}(t)$ on $[0,q]$, for a given $0\ll q <1$, and we want to predict $\{y_{n}(t): t\in (q, 1]\}$.
\end{itemize}
In both scenarios, we refer to the most recently observed functional datum, either $\{y_n(t): t\in [0,1] \}$ or $\{y_n(t): t\in [0,q] \}$, as the \emph{focal curve}. The latter will be denoted by $f$ and its observation domain by $D_f$, regardless of whether it is $[0,1]$ or $[0,q]$. For each $i$, $1\leq  i \leq n-1$, we consider the restriction of $y_i$ to $D_f$, this is the curve with graph equal to $ \{(t,y_i(t)): t\in D_f\}$. We denote this restriction by $y_{i|f}$ and we refer to ${\cal Y}_n = \{y_{1|f},\dots y_{n-1|f}\}$ as the \emph{past-focal-curves sample}. Note that in one-step-ahead  forecasting $y_{i|f}\equiv y_{i}$ .

Denote by $p$ the curve or curve segment to predict, this is $\{y_{n+1}(t): t\in [0,1] \}$ or $\{y_n(t): t\in (q,1] \}$. The domain of $p$, either $[0,1]$ or  $(q,1] $, is denoted by $D_p$. Then, for each $ y_{i|f}  \in {\cal Y} _n$, we consider the \emph{projection}
\begin{equation*}
{\cal P}y_{i|f} =
\begin{cases}
\{y_{i+1}(t), t\in [0,1]\}, &\mbox{in one-step-ahead  forecasting.}\\
\{y_{i}(t), t\in (q,1]\}, &\mbox{in dynamic updating.}\\
\end{cases}
\end{equation*}
Note that ${\cal P}f = p$ \bluetext{and that, although not explored it in this article, $h$-step-ahead projection could be achieved by ${\cal P}y_{i|f} = \{y_{i+h}(t), t\in [0,1]\}$}.

Many FTS regularly repeat dependency patterns between past focal curves and their projections. The change in shape and magnitude of the curves over time of some FTS can be visualized from the rainbow plots of \citet{Hyndman2010} and other related plots \citep{Shang2019}.

In dynamic updating, we observe that $y_{i|f} $ and $ {\cal P}y_{i|f}$ are a partition of $y_i$ in two curve segments. The dependence patterns between past focal curves and their projections can be observed from groups of curves with similar colors in the rainbow plots already mentioned. Similar curves may be distant in time due to seasons and cycles, making specific patterns be repeated over time. It is not necessarily an easy task to visualize dependency patterns between focal curves and their projections; it depends on the complexity of the FTS and the curves' shapes, and the number of curves involved. Nonetheless, the cyclic nature of many FTS suggests that there may be statistical regularity between the morphology of focal curves and the morphology of their projections. The core idea of the methods discussed here is based on the existence of this kind of regularity. Thus, if $f$ is surrounded by past focal curves that capture both its shape and magnitude, we could predict $p$ from these past focal curves' projections. Of course, when the FTS has a strong global trend, the future values are always out of the historical data magnitude. Therefore, the projection methods discussed here are not suitable without detrending in such cases. 

\subsection{The focal-curve envelope} \label{sec:curvesSelection}

The straightforward approach for surrounding $f$ with past focal curves is by using KNN. However, as we already commented, \bluetext{$f$ does not have to be necessarily centred in the band delimited by its $k$-nearest curves}. Even $f$ may be completely outside of this band. To produce bands where $f$ is central, which is natural for band estimation, we use the concept of functional depth. In particular, we use the modified band depth of \citet{pintadoromo2009} with bands formed by two curves. This is a widely used depth measure that allows a million curves to be ranked in only tenths of seconds \citep{FastBD} making our algorithm efficient even for large data sets. Strictly speaking, for any $\mathcal{J}\subset {\cal Y}_n$ with $|\mathcal{J}| \geq 2$ and $y\in \mathcal{J}\cup \{f\}$, we consider
\begin{equation*}
D(y,\mathcal{J})=  \frac{1}{2}{|\mathcal{J}|+1 \choose 2}^{-1}\sum_{x,z\in \mathcal{J}\cup \{f\}}  \lambda \left(\left\lbrace t\in D_f: \min (x(t),z(t)) \leq y(t) \leq \max (x(t),z(t)) \right\rbrace \right),
\end{equation*}
$\lambda(\{t\in D_f:A(t)\})$ being the proportion of time that $A(t)$ is true on $D_f$. The larger $D(y,{\cal J})$, the deeper will be $y$ in ${\cal J}\cup \{f\}$. Denote by ${\cal J}_k$ the $k$ deepest curves of ${\cal J}$. We remark that $f\notin {\cal J}_k$ although $f$ may be the deepest curve of ${\cal J}\cup \{f\}$.

In line with  \citet{sungenton2011}, who consider central regions based on the modified band depth for visualizing the spread of the deepest curves, we consider the \emph{Focal Central Region} (FCR) delimited by ${\cal J}_k$. This is the region
\begin{equation*}\label{eq:centralregion}
R({\cal J}_k) = \left\{(t,y(t)): t\in D_f, \min_{x\in {\cal J}_k}x(t) \leq y(t) \leq \max_{x\in {\cal J}_k} x(t)\right\}.
\end{equation*}

So that $R({\cal J}_k)$ can capture the shape and magnitude of $f$, $R({\cal J}_k)$ must be roughly centered at $f$  and its mean width should be small. Having these objectives in mind, we look for a set $ {\cal J} $ of past focal curves, as large as possible in order of not restricting the range of $k$, such that:
\begin{enumerate} 
	\item[1)] $f$ is deep in ${\cal J}\cup \{f\}$, the deepest if possible.
	\item[2)] $f$ is enveloped by ${\cal J}$ as much as possible. Here, we say $f$ is \emph{more enveloped} by ${\cal J}$ than by ${\cal J}'$ if and only if $$\lambda\left( \left\{t: \min_{y\in {\cal J}}y(t) \leq f(t) \leq \max_{y\in {\cal J}} y(t) \right\} \right) > \lambda\left(\left\{t: \min_{y\in {\cal J}'}y(t) \leq f(t) \leq \max_{y\in {\cal J}'} y(t) \right\}\right).$$
	\item[3)] $f$ is surrounded by near curves of ${\cal J}$, as many as possible.
\end{enumerate}
Algorithm~\ref{alg:Alg1} provides a set of past focal curves with the three features above that we call the \emph{focal-curve envelope} and denote by $ {\cal J}$ from now on. In a nutshell, the algorithm iteratively selects as many past curves as possible. From the nearest to the farthest to $f$, we obtain an envelope of $f$ with an increasing depth.

\algdef{SE}[SUBALG]{Indent}{EndIndent}{}{\algorithmicend\ }%
\algtext*{Indent}
\algtext*{EndIndent}
\begin{algorithm}
	\caption{Input: ${\cal Y}_n, f$. Output: $\mathcal{J}$}
	\label{alg:Alg1}
	\begin{algorithmic}[]
		\small
		\State \textbf{Initialize} ${\cal Y} = {\cal Y}_n$, ${\cal J} = \emptyset$, $D(f,{\cal J})=0$
		\While{ size of ${\cal Y} \geq 2 $ }
		\State Let $y'$ be the nearest curve to $f$ from $ {\cal Y}$ and ${\cal N} = \{y'\}$
		\For{ $y\in {\cal Y}\setminus \{y'\}$, from the nearest curve to the farthest from $f$,}
		\If{$f$ is more enveloped by  ${\cal N}\cup \{y\}$ than by ${\cal N}$}
		\State ${\cal N} = {\cal N} \cup \{y\}$ 
		\EndIf
		\EndFor
		\If{$D(f, {\cal J}\cup {\cal N}) \geq D(f,{\cal J})$}
		\State ${\cal J} = {\cal J} \cup  {\cal N}$ 
		\EndIf
		\State ${\cal Y} = {\cal Y} \setminus  {\cal N}$
		\EndWhile
	\end{algorithmic}
\end{algorithm}

\begin{figure}[!htbp]
	\centering
	\includegraphics[width=1\textwidth]{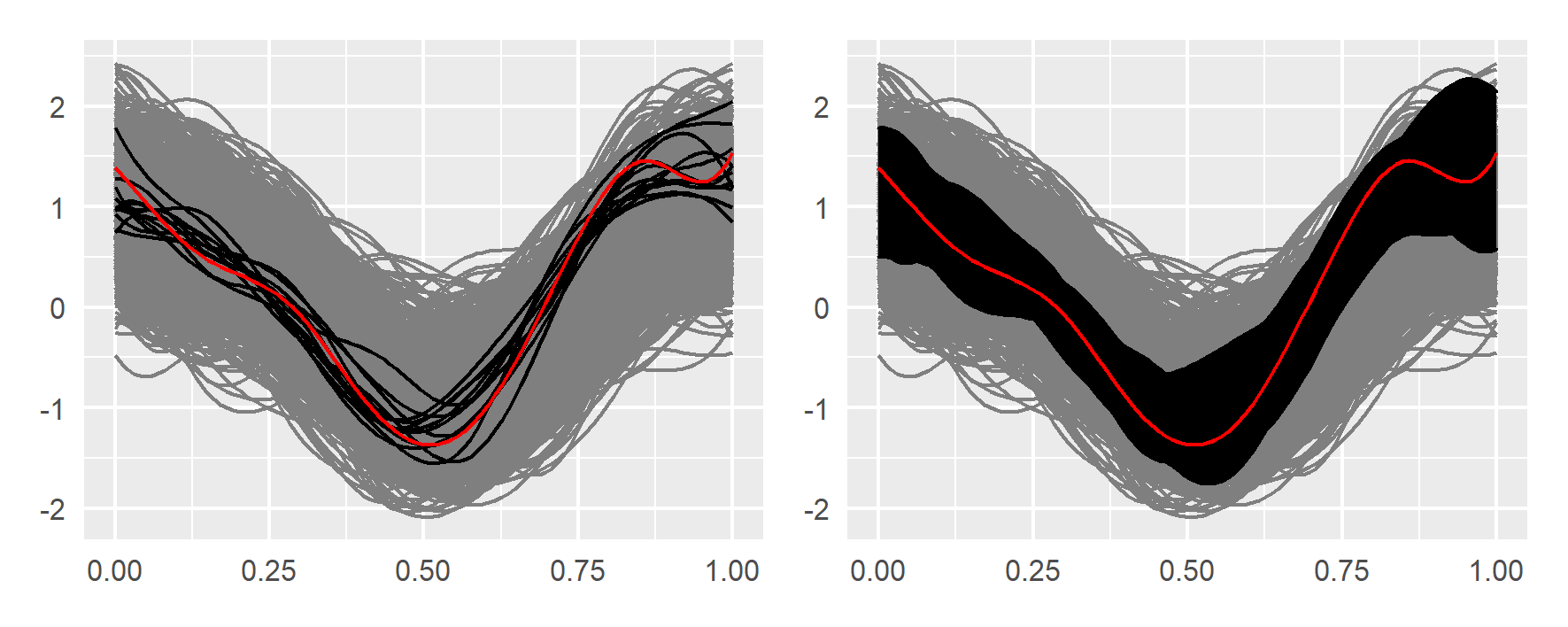}
	\includegraphics[width=1\textwidth]{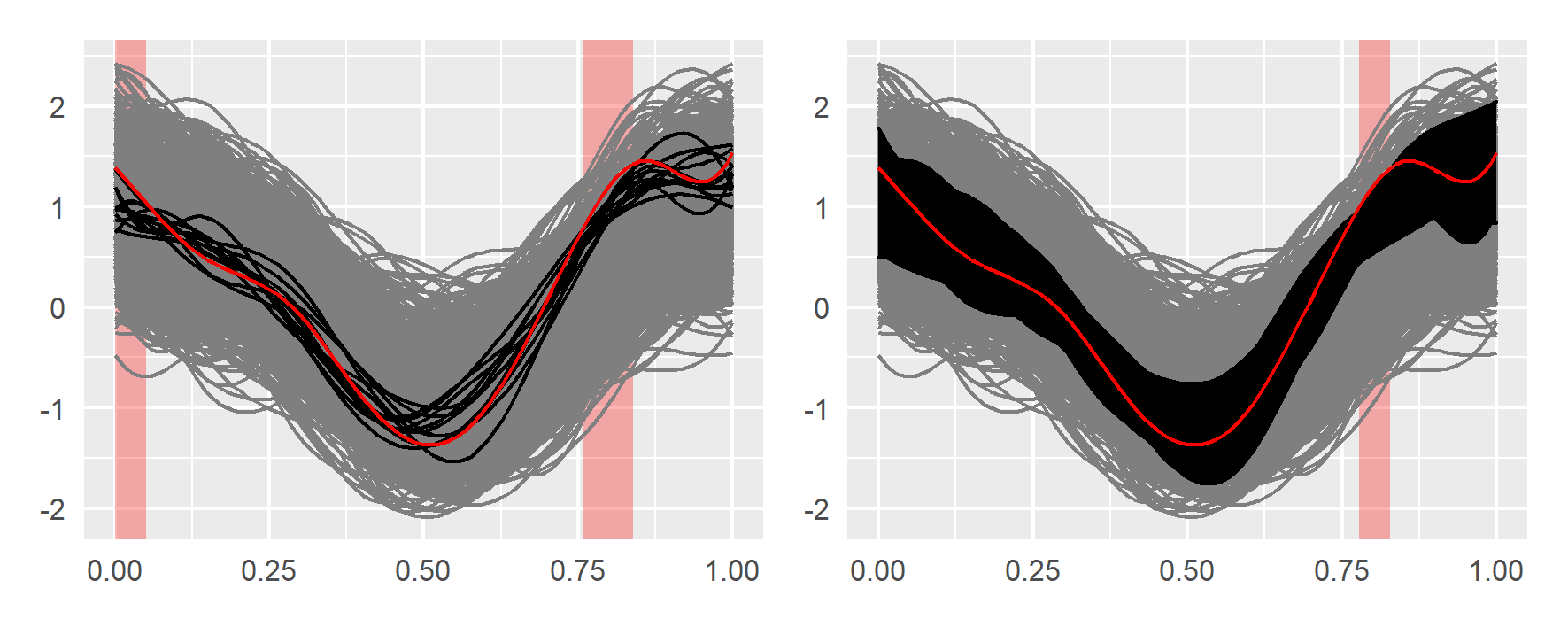}
	\caption{Top panels: focal curves (in red), the twelve curves selected on the first iteration of Algorithm~\ref{alg:Alg1} (left panel) and band delimited by the focal-curve envelope from the 46 curves selected by the algorithm (right panel). Bottoms panels: The 12-nearest curves and the band delimited by the 46-nearest curves. The intervals where the nearest curves do not envelope the focal are shaded in red. Algorithm~\ref{alg:Alg1} selects near curves that envelope the focal until making it been the deepest one. The nearest curves do not necessarily envelope the focal curve nor making it been the deepest one.}
	\label{fig:1}
\end{figure}

As an illustration of how Algorithm~\ref{alg:Alg1} works, see top panels of Figure~\ref{fig:1}, where 100 curves from a simulated FTS are considered. The left panels show the twelve curves selected on the first iteration of Algorithm~\ref{alg:Alg1} (top panel) and the 12-nearest neighbors (bottom panel). For measuring nearness, hereafter, we use $L_2$ distance. Besides this is the most widely used distance in the KNN context, we base our choice by looking for a small mean squared error on the prediction method discussed in Subsection~\ref{subset:prediction}. We emphasize the intervals where the nearest curves do not envelope the focal curve. Right panels show the band delimited by the focal-curve envelope obtained after five iterations and formed by forty-six curves and the corresponding band from the 46-nearest curves. As a general rule, Algorithm~\ref{alg:Alg1} selects near curves that are enveloping the focal \bluetext{until it is the deepest one. In contrast, nearest neighbors neither necessarily envelope the focal curve nor make it the deepest one.}

\subsection{Prediction}\label{subset:prediction}

We make predictions by projecting the $k$-nearest curves to the focal and projecting the curves that belong to the focal-curve envelope. We compute weighted means of these projections for punctual prediction, assigning more weights to the curves closer to the focal. Following the literature on KNN \citep{Martinez2019, kara2017}, and simplex projections or S-map \citep{perretti2013}, we consider two types of weights: On one hand, weights inversely proportional to the square distance to the focal, which we identify in the sequel by $1/d$ weights. On the other hand, exponential weights. Specifically, let $f_1, \dots, f_k$ be the $k$-nearest curves to $f$, $d_i = \lVert f_i -f\rVert ^2$, and more general $d_y = \lVert y -f\rVert ^2$, let ${\cal J}$ be the focal-curve envelope of $f$ and $\theta>0$, what we do is to consider the following estimators of $p={\cal P}f$:
\begin{itemize}
	\item functional KNN projections (fKNN) with $1/d$ and exponential weights
	\begin{equation}\label{KNNestimator}
	\hat{p}_k = \frac{\sum_{i=1}^k  \frac{1}{d_i} \ {\cal P}f_i} {\sum_{i=1}^k  \frac{1}{d_i }},\ \ \mbox{and} \ \ \hat{p}_{k,\theta}  = \frac{\sum_{i=1}^k  e^{-\theta d_i/d_1}  {\cal P}y}{\sum_{i=1}^k  e^{-\theta d_i/d_1}}
	\end{equation}
	\item Envelope projections (EP) with $1/d$ and exponential weights
	\begin{equation}\label{eq:pointestimator}
	\hat{p} = \frac{\sum_{y\in {\cal J}}  \frac{1}{d_i } \ {\cal P}f_i} {\sum_{y\in {\cal J}}  \frac{1}{d_i }},\ \ \mbox{and} \ \ \hat{p}_{\theta}  = \frac{\sum_{y\in {\cal J}}  e^{-\theta d_y/d_1}  {\cal P}y}{\sum_{y\in {\cal J}}  e^{-\theta d_y/d_1}}
	\end{equation}
\end{itemize}

For fixed $\theta$, $k$ in~\eqref{KNNestimator} will be chosen by minimizing the prediction Mean Square Error (MSE) on a rolling forecasting origin procedure, as it is customary in the literature of KNN for time series. We denote by $k^*$ the value that minimizes the MSE. 
Similarly, we can select $\theta$ in~\eqref{eq:pointestimator}. Let $\theta^*$ be the resulting value of minimizing MSE by rolling origin. We discard selection of optimal pairs $(\theta,k)$ for $\hat{p}_{k,\theta} $ in~\eqref{KNNestimator} by a high computational cost involved in this two-dimensional optimization problem. To compare the methods, on one hand we compute $\hat{p}_k^*$ and $\hat{p}$. On the other hand, we fix $\theta = 1$ and compute $\hat{p}_{k^*,\theta = 1}$ and $\hat{p}_{\theta = 1}$. In addition, we compute $\hat{p}_{\theta^*}$.

For band prediction, we consider the region delimited by the projections of the $k$ deepest curves of $ {\cal J}$. This is the band
\begin{equation*}\label{eq:band}
{\cal P}R({\cal J}_{k}) = \left\{(t,y(t)): t\in D_p, \min_{x\in {\cal J}_{k}} {\cal P}x(t) \leq y(t) \leq \max_{x\in {\cal J}_{k}} {\cal P}x(t)\right\}.
\end{equation*}
This band prediction corresponds to a central functional region such as those defined by \cite{sungenton2011}. As with the fKNN projections, $k$ is chosen by minimizing a rolling forecasting origin procedure's cost function. For that, we consider the Winkler score. This is a measure that penalizes both bandwidth and non-coverage for a given $\alpha \in (0,1)$, penalizing even more heavily when observations are remarkably outside the band the confidence levels are higher \citep{Winkler1972, Gneiting2007}. Thus, for a given $(1-\alpha)$ mean coverage, i.e., the proportion of time that $p$ is in the band, $k$ is chosen by minimizing the score. Mimicking this procedure, we also consider the region delimited by the projection of the $k$ nearest curves for band prediction based on fKNN.

\begin{figure}[!htbp]
	\centering
	\includegraphics[width=1\textwidth]{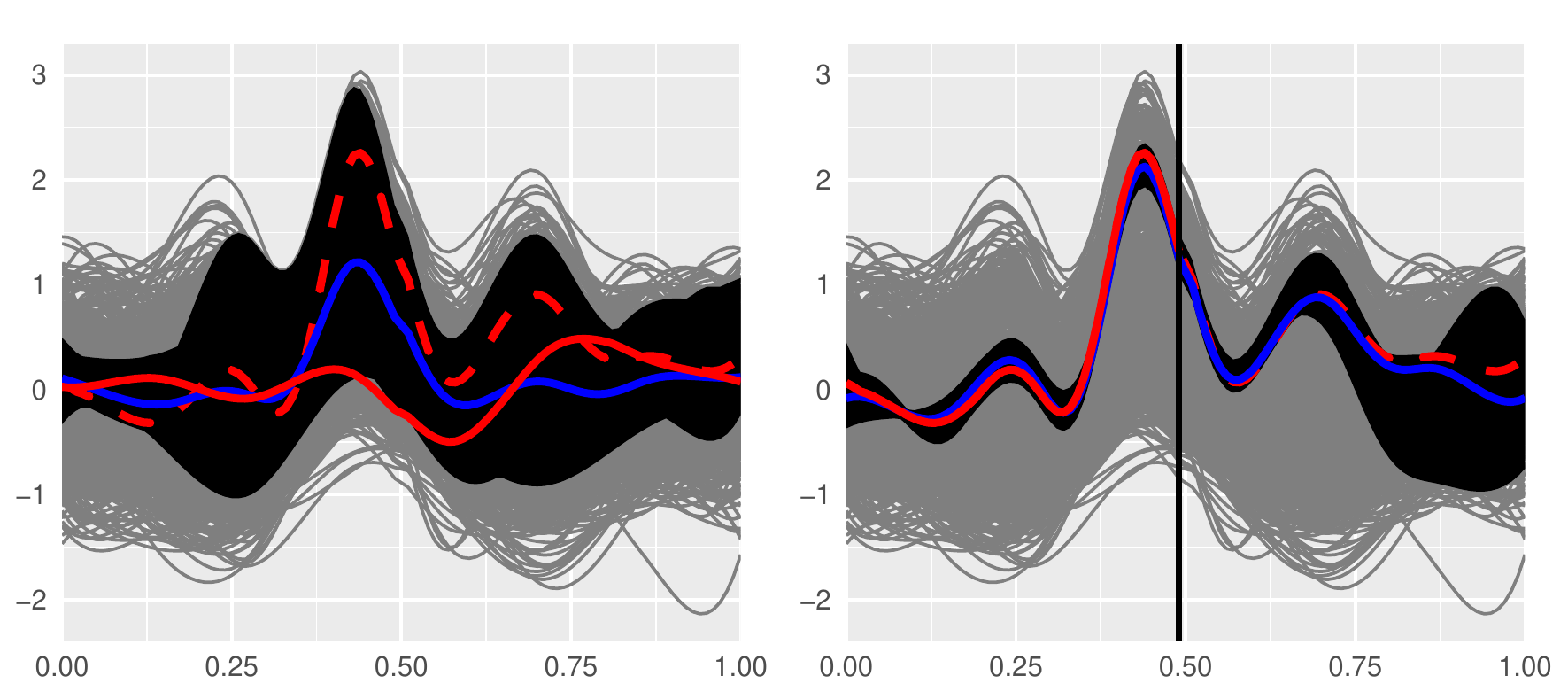}
	\caption{1000 curves from a simulated FTS (grey). Focal curve $f$ in solid red, curve to predict $p$ in dashed red, point prediction $\hat{p}_1$ in blue and prediction band $ {\cal P}R({\cal J}_{10})$ in black. Left panel: One-step-ahead prediction. Right panel: Dynamic updating from $0.5$ to $1$.}\label{fig:2}
\end{figure}

Figure~\ref{fig:2} shows an example of point and band prediction based on 1000 curves from a simulated FTS corresponding to the model discussed in Section~\ref{sec:PCP} with $\mu = 0$. Both one-step-ahead forecasting and dynamic updating provided by EP with $\hat{p}_{\theta}$ are illustrated.

\section{Results}\label{sec:results}

For one-period-ahead forecasting, we compare fKNN and EP estimators with benchmark predictions for stationary FTS. In particular, we implement a variant of the method proposed by \citet{hyndman2007} and \citet{Hyndman2009}, who suggest functional prediction based on modeling the functional principal component scores as independent time series. Following \citet{Aue2015}, we will consider the first $d$ functional scores as a $d$-variate time series since the score vectors could have non-diagonal autocorrelations. The number of scores $d$ is chosen as the smallest integer such that the first $d$ principal components explain at least 80\% of the variance of the data. Then, we fit a $d$-dimensional autoregressive model \citep{lutkepohl2006} to the scores' time series. This method is termed FPCF. For validating FPCF as a benchmark method, we repeated the simulation study conducted by \citet{Aue2015} and obtained competitive results with the best of their methods (see Appendix~\ref{sec:appendixFAR}). 

For dynamic updating, we compare fKNN, and EP with the method of \cite{Kraus2015} for predicting partially observed functional data. We also tested the well-known method of \citet{Yao2005} but \citet{Kraus2015} achieved better results in our experiments. When the FTS is constructed by slicing a periodic univariate time series, we also consider the Block-Moving (BM) approach by \citet{hyndmanshang2011}. The BM approach is designed to re-arrange the time series so that all the FTS are ultimately observed and, then, to apply FPCF is possible. The literature devoted to dynamic updating has successfully proposed methodologies based on ordinary least squares (OLS) regression, ridge regression (RR), penalized least squares (PLS) regression, or functional linear regression \cite{hyndmanshang2011, Shang2017, Shang2018}. However, we do not compare with them because we deal with large data sets where these methods might be computationally inefficient. In other circumstances, they certainly are the benchmark of the field.

Finally, to understand the scale of how superior a method is with respect to the other, we also report about the two most basic estimators. They are the historical mean and the naive prediction. This is just the most recent observed datum corresponding to the prediction range. In our notation, the naive predictor of ${\cal P}y_n$ is ${\cal P}y_{n-1}$.

The implementation of the Envelope Projection method is available at \url{https://github.com/aefdz/nnFTS}. The benchmark methods for Functional Time Series are available in the \texttt{ftsa} R package by \citet{ftsaR}. Finally, the method for estimating partially observed data was implemented by \cite{Kraus2015} with the code available from \url{https://www.davidkraus.net/?page_id=349}.

\subsection{Stationary FTS with random shocks}\label{sec:PCP}

\begin{figure}[!htbp]
	\begin{center}
		\includegraphics[width=1\textwidth]{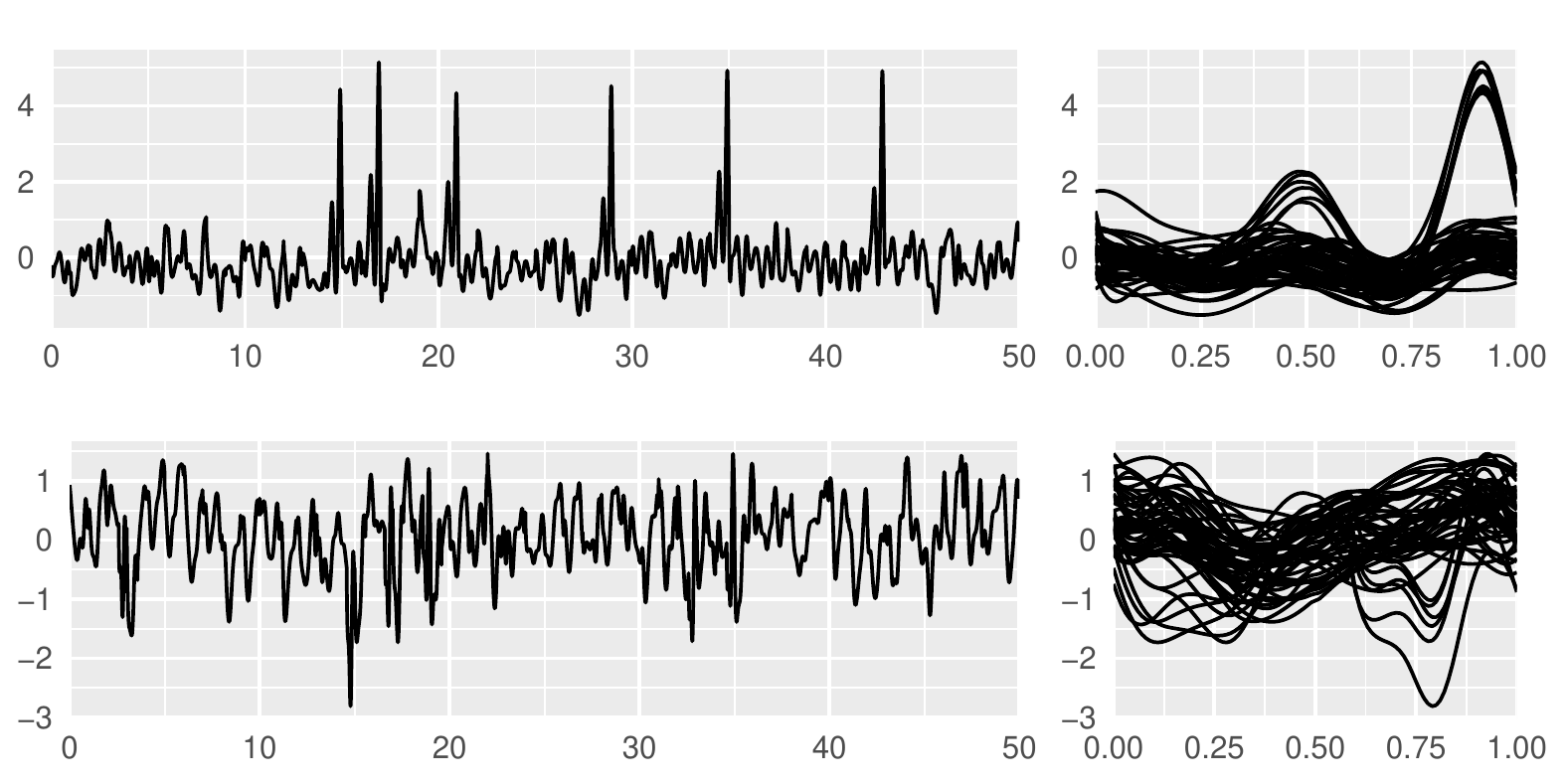}
		\caption{Two random trajectories of $Y_{0.2}$ (left panel) and their corresponding FTS (right panel).}
		\label{fig:3}
	\end{center}
\end{figure}

To simulate FTS with the features discussed at the end of Section~\ref{sec:basicNotions}, we consider slices of a scalar process with complex periodic rhythm \bluetext{that artificially corrupt with shocks}. \bluetext{This process is denoted \bluetext{$Y_\mu$}, being $\mu \in [0,1)$ the expected proportion of affected periods}. The process is the sum of three independent components corresponding to:
\begin{enumerate}
	\item A noise $\{X(t): t\geq 0\}$, modeled by a zero-mean stationary Gaussian process.
	\item A random periodic function $f:[0,+\infty)\rightarrow \mathbb{R}$. 
	\item A random shock process $\{S_\mu(t): t\geq 0\}$ that occasionally affects two contiguous periods of $f$. \bluetext{For $\mu = 0$, $S_{\mu}=0$}.
\end{enumerate}
See Appendix~\ref{sec:appendixA} for more details about these processes. Thus $Y_\mu = X + f + S_\mu$ and the considered FTS is $y_i(t) = Y_\mu(t+i-1)$, $0\leq t \leq 1$, $i\in \mathbb{N}$. If $\mu =0$, there are no shocks and the corresponding FTS is stationary. With increasing $\mu$, we explore different non-stationary regimes. Figure~\ref{fig:3} shows two random trajectories of $Y_{0.2}$ and their respective FTS.

We performed 300 realizations of $Y_\mu$, 100 for each level of contamination considered. They were $\mu = 0.0, 0.2$ and $0.4$. For each realization, we considered $1000$ periods and, from the corresponding $1000$ slices, the exercise of one-step-ahead prediction and dynamic updating on the last $100$ slices. For dynamic updating, we chose $q = 0.5$, corresponding to half slice. The resulting MSEs are shown in Table~\ref{tab:shocks}. In summary, we found the proposed methods were superior to the benchmark method. They were even better at higher contamination levels, and fKNN is often superior to EP.

\begin{table}[!htbp]
	\setlength{\tabcolsep}{8.6pt}
	\caption{Mean Squared Errors for one-step-ahead forecasting and dynamic updating from a stationary FTS ($\mu = 0$) contaminated with random shocks at two levels of contamination ($\mu = 0.2$ and $\mu = 0.4$). The observed minimum MSE among the considered methods is remarked in bold.}\label{tab:shocks}
	\begin{center}
		\begin{subtable}{1\textwidth}
			\begin{tabular}{@{}lccccccccc@{}}
				\toprule
				\multicolumn{10}{l}{\hspace{-.1in}{One-step-ahead prediction}} \\
				\cmidrule{1-10} 
				%\addlinespace[0.1cm]
				& & & \multicolumn{2}{c}{fKNN} & &\multicolumn{3}{c}{EP} &  \\
				\cmidrule{4-5} \cmidrule{7-9}  
				%\addlinespace[0.1cm]
				& Average & Naive & $\hat{p}_{k^*}$ & $\hat{p}_{k^*,\theta=1}$ & & $\hat{p}$ & $\hat{p}_{\theta=1}$ & $\hat{p}_{\theta^*}$ & FPCF \\ 
				\cmidrule{1-10} 
				\addlinespace[0.1cm]
				$ \mu = 0.0$ & 0.192 & 0.385 & 0.1880 & \textbf{0.1879} & & 0.192 & 0.199 & 0.192 & 0.190  \\ 
				\addlinespace[0.1cm]
				$ \mu = 0.2$ & 2.465 & 5.380 & 1.5288 & \textbf{1.5153} & & 1.646 & 1.679 & 1.594 & 1.662 \\
				\addlinespace[0.1cm]
				$ \mu = 0.4$ & 4.513 & 11.35 & \textbf{2.5893} & 2.5974 & & 2.912 & 2.832 & 2.744 & 2.767 \\ 
				\addlinespace[0.3cm]
			\end{tabular}
		\end{subtable}
		
		\begin{subtable}{1\textwidth}
			\begin{tabular}{@{}lcccccccccc@{}}
				\toprule
				\multicolumn{11}{l}{\hspace{-.1in}{Dynamic Updating}} \\ 
				\midrule
				&  &  & \multicolumn{2}{c}{fKNN} & & \multicolumn{3}{c}{EPU} & & \\
				\cmidrule{4-5} \cmidrule{7-9}  
				& Average & Naive & $\hat{p}_{k^*}$ & $\hat{p}_{k^*,\theta=1}$ & & $\hat{p}$ & $\hat{p}_{\theta=1}$ & $\hat{p}_{\theta^*}$ & BM & Kraus \\ 
				\midrule
				\addlinespace[0.1cm]
				$ \mu = 0.0$ & 0.192 & 0.382  & 0.1582 & 0.1582 & & 0.160 & 0.172 & \textbf{0.157} & 0.182 & 0.166 \\
				\addlinespace[0.1cm]
				$ \mu = 0.2$ & 2.527 & 5.520 & 1.2109 & \textbf{1.1920} & & 1.669 & 1.421 & 1.239 & 2.173 & 1.599 \\
				\addlinespace[0.1cm]
				$ \mu = 0.4$ & 4.783 & 12.126 & 1.4078  & \textbf{1.3805} & & 2.278 & 1.640 & 1.418 & 3.756 & 2.750 \\
				\bottomrule
			\end{tabular}
		\end{subtable}
	\end{center}
\end{table}

\subsection{Spanish electricity demand}\label{sec:electricity}

Functional methods for electricity demand forecasting have been widely used \citep{Vilar2012, Antoch2010, Paparoditis2013, Cho2013, Shang2013, Aneiros2013, aneiros2016, Vilar2018, Rana2018}. These methods often consider particular features of the data, such as weekly seasonality, the effect of holidays, and outliers' presence. Thus, researchers process weekdays separately from Saturdays and Sundays and apply methods for replacing outliers. In addition to this, \citet{Rana2018} use additional data from functional exogenous covariates in their Functional Additive Models (FAM) for next-day forecasting. Certainly, they provide the benchmark predictions in functional Spanish electricity demand forecasting.

We consider the challenge of predicting the daily curves of a whole year (2018) without considering any specific peculiarity of the electricity demand by using raw data. For dynamic updating, we considered the practical case of half-day ahead prediction. Our study is based on electricity demand at every 10-minute interval, from January 1, 2014, to December 31, 2018. The data set is available at \url{http://www.ree.es/es/}. Following \citet{Rana2018}, we compute the mean absolute percentage error (MAPE) in addition to the MSE. Viewing the weekly seasonality of electricity demand and the naive prediction corresponding to yesterday for predicting today, we also consider the naive prediction corresponding to the same day one week ago, identified below as seasonal naive (Naive$'$ in the tables). The results are shown in Table~\ref{tab:electricity}.

\begin{table}[!htbp]
	\centering
	\caption{Forecasting accuracy for daily electricity demand during 2018. The performance is measure by MSE, MSE/MSE$_{min}$, being the MSE divided by the minimum MSE among the different methods and MAPE. The minimum value of each row is remarked in bold.} \label{tab:electricity}
	\begin{subtable}{\textwidth}
		\resizebox{\textwidth}{!}{
			\begin{tabular}{@{}lcccccccccc@{}}
				\toprule
				\multicolumn{10}{l}{One-step-ahead prediction} \\
				\midrule
				&  &  & & \multicolumn{2}{c}{fKNN} & & \multicolumn{3}{c}{EP} & \\
				\cmidrule{5-6} \cmidrule{8-10}  
				& Average  & Naive & Naive' & $\hat{p}_{k^*}$ & $\hat{p}_{k^*,\theta=1}$ & & $\hat{p}$ & $\hat{p}_{\theta=1}$ & $\hat{p}_{\theta^*}$  & FPCF    \\ 
				\cline{2-11}
				\addlinespace[0.1cm]
				MSE  & 10139732 & 8011267 & 4448292 & 1537719   & 1530080 & & 2144232 & 1440256 & \textbf{1416428} & 5771944 \\
				\addlinespace[0.1cm]
				$\frac{\text{MSE}}{\text{MSE}_{\min}}$ & 7.158 & 5.655 & 3.140   & 1.086     & 1.080 &        & 1.514   & 1.017 & \textbf{1.000}   & 4.075   \\
				\addlinespace[0.1cm]
				MAPE & 8.857 & 6.383 & 4.647 & 2.57953 & 2.59538 & & 3.495 & 2.666 & \textbf{2.542} & 6.583 \\
				\addlinespace[0.3cm]
			\end{tabular}
		}
	\end{subtable}
	\begin{subtable}{\textwidth}
		\resizebox{\textwidth}{!}{
			\begin{tabular}{@{}lccccccccccc@{}}
				\toprule
				\multicolumn{11}{l}{Dynamic Updating} \\
				\midrule
				&  &  & & \multicolumn{2}{c}{fKNN} & & \multicolumn{3}{c}{EP} & \\
				\cmidrule{5-6} \cmidrule{8-10}  
				& Average  & Naive & Naive' & $\hat{p}_{k^*}$ & $\hat{p}_{k^*,\theta=1}$ & & $\hat{p}$ & $\hat{p}_{\theta=1}$ & $\hat{p}_{\theta^*}$  & BM & KRAUS  \\ \cline{2-12}
				\addlinespace[0.1cm]
				MSE & 11545143 & 7805649 & 5245784 & 433620  & 426642 & & 534009  & 414877 & \textbf{413508}  & 2823846 & 512749.9 \\
				\addlinespace[0.1cm]
				$\frac{\text{MSE}}{\text{MSE}_{\min}}$ & 27.92    & 18.877  & 12.687  & 1.049   & 1.032 & & 1.291   & 1.003        & \textbf{1.000}   & 6.829   & 1.240    \\
				\addlinespace[0.1cm]
				MAPE & 9.061 & 6.183   & 5.035   & 1.55566 & 1.55052 & & 1.79564 & 1.54113 & \textbf{1.53637} & 4.43493 & 1.7558  \\
				\bottomrule
			\end{tabular}%
		}
	\end{subtable}
\end{table}

\bluetext{From this table, we remark that FPCF is slightly inferior to the naive method and significantly inferior to seasonal naive. However, the projection methods were superior to the benchmark competitors for both one-step-ahead forecasting and dynamic updating, and EP slightly superior to fKNN.}

\bluetext{Although they are different studies, MAPE of \citet{Rana2018} (in the last four rows of the last column of their Table 1) might reference the expected magnitude of the forecasting performance. The naive method in both studies produces the same MAPE (6.383 vs. 6.39). However, we remark that MAPE computed by \citet{Rana2018} corresponds to 2012 prediction instead of 2018, using demand each hour, instead of every 10 minutes, and therefore conclusions might be taken with caution. From this extrapolation, FPCF in Table~\ref{tab:electricity} seems to be by far inferior to FAM. This points out the convenience of removing outliers and taking into account the data's weekly seasonality to improve predictions as suggested by \citet{Rana2018}. Finally, MAPE of fKNN and EP turned out to be by far smaller than those of FAM, despite being based on data without preprocessed, without splitting the data into Weekdays, Saturdays, and Sundays, and without covariates.}

Notably, \citet{Rana2018} also reports coverage and mean Winkler scores of prediction bands, \bluetext{providing an approximate order of magnitude of the results to compare them} with the prediction bands based on projection methods introduced at the end of Subsection~\ref{subset:prediction}. Table~\ref{tab:4} displays the scores resulting from projection methods for $\alpha = 0.05, 0.10$ and $0.20$. The first noteworthy result is that the mean coverages obtained are often above $(1-\alpha)$. Second, the mean Winkler scores are often smaller than that those previously reported.

\begin{table}[!htbp]
	\caption{Coverage and mean Winkler score of prediction bands for electricity demand and NOx emissions.} 
	\label{tab:4}
	\centering
	\resizebox{\textwidth}{!}{
		\begin{tabular}{@{}lccccccccccccc@{}}
			\toprule
			& & \multicolumn{5}{c}{Electricity Demand} & & \multicolumn{5}{c}{NoX Emission} \\			
			\cmidrule{3-7}  \cmidrule{9-13}			
			& & \multicolumn{2}{c}{One-day-ahead} & & \multicolumn{2}{c}{Dynamic Updating} & & \multicolumn{2}{c}{One-day-ahead} & & \multicolumn{2}{c}{Dynamic Updating} \\			
			\cmidrule{3-4} \cmidrule{6-7}  \cmidrule{9-10}  \cmidrule{12-13}			
			$\alpha$ & & EPF & fKNN & & EPU & fKNN & & EPF & fKNN & & EPU & fKNN  \\ 
			\hline
			$0.05$  &  Coverage & 0.96 & 0.96 & & 0.96 & 0.93 & & 0.94 & 0.94  & & 0.96 & 0.98 \\ 
			& (width)  & 6340.22 & 5581.59 & & 3909.38 & 6871.60 &  & 88.64 & 81.07 & & 99.56 & 95.59 \\ 
			& Winkler & 7245.76 & 6200.67 & & 4391.11 & 10326.45 & & 103.94 & 94.95 & & 108.16 & 97.85 \\ 
			\addlinespace[0.3cm]
			$0.10$  &  Coverage & 0.93 & 0.92 & & 0.92 & 0.87 & & 0.89 & 0.89 & & 0.89 & 0.95 \\ 
			& (width) & 5272.801 & 4477.16 & & 3126.253 & 5881.41 & & 71.16 & 68.91 & & 60.41 & 79.47 \\ 
			&  Winkler & 6005.07 & 5097.22 & & 3672.03 & 9969.236 & & 87.23 & 84.90 & & 71.70 & 83.81 \\ 
			\addlinespace[0.3cm]
			$0.20$  &  Coverage & 0.83 & 0.84 & & 0.82 & 0.80 & & 0.80 & 0.79 & & 0.80 & 0.87 \\ 
			& (width) &  4228.828 & 3355.45 & & 2457.981 & 4794.38 & &   56.93 & 54.54 & & 47.43 & 64.04 \\ 
			&  Winkler & 5390.77 & 4185.80 & & 3162.14 & 8344.702 & & 75.13 & 75.06 & & 60.34 & 71.79  \\ 
			\bottomrule
		\end{tabular}
	}
\end{table}

\begin{figure}[!htbp]
	\begin{center}
		\includegraphics[width=1\textwidth]{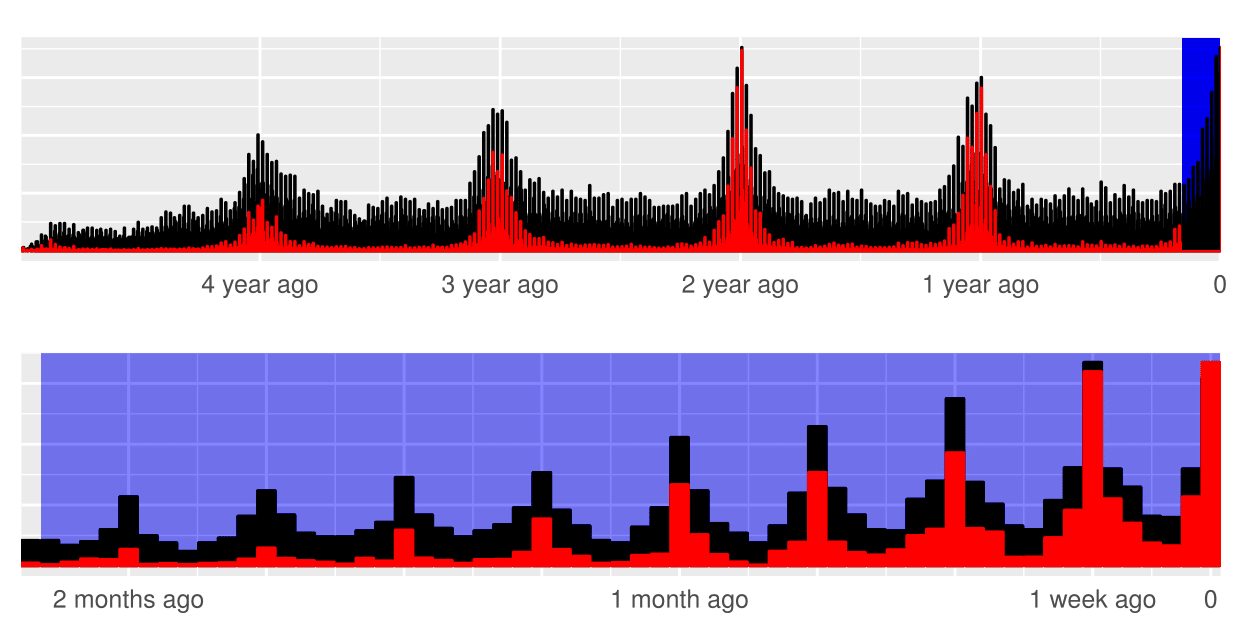}
		\caption{Black bars: frequency distribution of the number of days between predicted day and the ones used in the EP prediction. Red bars: the normalized average of exponential weights used in the EP prediction. Bottom panel: zoom of the blue shaded area in the top panel.}
		\label{fig:4}
	\end{center}
\end{figure}

Also, we want to remark that the projection methods take into account the weekly and yearly periodicity of the data without any prior information. For illustrating this, we plot the frequency distribution of the number of days between predicted days and days used in the EP prediction in Figure~\ref{fig:4}. The exponential weights properly averaged and scaled are also plotted in the same figure. This figure reveals the most relevant past information to forecast one-period-ahead and the electricity demand's seasonal structure. In particular, we observe that the most recent day contributes the most to the point forecast, followed by the same weekday but one week before (highest bars at the bottom panel). This weekly contribution vanishes until we approach the values of one year ago (highest bars at the top panel). For example, if we aim to forecast the demand for a given Wednesday, then the envelope would consider data from the last Tuesday, past Wednesdays, and same calendar days of other years in the same season, in order of importance.

\subsection{NOx emissions in Madrid city center}\label{sec:NOX}

Air quality is nowadays in the spotlight due to its impact on human health \citep{medAir}. Many cities gather almost continuous data of different air quality measures, such as nitrogen monoxide (NOx). The Cityhall of Madrid has recorded data since $2001$ at different city points and has made them publicly available at \url{https://datos.madrid.es/portal/site/egob/}. 
In this section, we analyze the data from a station located in Plaza de Espa\~{n}a, Madrid downtown, because it has been functioning uninterrupted since the very beginning. By discarding days with fails on record, we end up with $6187$ daily functions from $2001$ to $2017$ observed each hour. 

Following \cite{ramsay2005}, we evaluate these functions every 15 minutes by applying a positive smoothing with $K=11$ Fourier basics functions (periodic data) and a $\lambda$ obtained from minimizing the generalized cross-validation coefficient. As we did with the Spanish electricity demand, we forecast the last recorded year (2017), predicting half a day for dynamic updating exercise. The results are also shown in Table~\ref{tab:NOX}.

\begin{table}[!htbp]
	\centering
	\caption{Forecast accuracy for NOx emissions during 2017. The performance is measure by MSE/MSE$_{min}$, being the MSE divided by the minimum MSE among the different methods. The minimum value of each row is remarked in bold.}\label{tab:NOX}
	\begin{subtable}{\textwidth}
		\resizebox{\textwidth}{!}{
			\begin{tabular}{@{}lcccccccccc@{}}
				\toprule
				\multicolumn{10}{l}{\hspace{-0.08in}{One-step-ahead}} \\
				\midrule
				& & & & \multicolumn{2}{c}{fKNN} & & \multicolumn{3}{c}{EP} & \\
				\cmidrule{5-6} \cmidrule{8-10}  
				& Average  & Naive & Naive' & $\hat{p}_{k^*}$ & $\hat{p}_{k^*,\theta=1}$ & & $\hat{p}$ & $\hat{p}_{\theta=1}$ & $\hat{p}_{\theta^*}$  & FPCF \\ 
				\cmidrule{2-11} 
				\addlinespace[0.1cm]
				MSE & 661.5623  & 582.1691 & 945.5869 & 386.1636  & 386.2248 & & 383.0795 & 387.2456 & \textbf{381.9375} & 502.6298 \\
				\addlinespace[0.1cm]
				$\frac{\text{MSE}}{\text{MSE}_{min}}$ & 1.732 & 1.524    & 2.476 & 1.011 & 1.011 & & 1.003    & 1.014        & \textbf{1.000}    & 1.316 \\
				\addlinespace[0.3cm]
			\end{tabular}
		}
	\end{subtable}
	\begin{subtable}{\textwidth}
		\resizebox{\textwidth}{!}{
			\begin{tabular}{@{}lcccccccccc@{}}
				\toprule
				\multicolumn{11}{l}{\hspace{-0.08in}{Dynamic Updating}} \\
				\midrule
				&  &  & & \multicolumn{2}{c}{fKNN} & & \multicolumn{3}{c}{EP} & \\
				\cmidrule{5-6} \cmidrule{8-10}  
				& Average  & Naive & Naive' & $\hat{p}_{k^*}$ & $\hat{p}_{k^*,\theta=1}$ & & $\hat{p}$ & $\hat{p}_{\theta=1}$ & $\hat{p}_{\theta^*}$  & KRAUS  \\ \cline{2-11}
				\addlinespace[0.1cm]
				MSE & 733.5805 & 512.6575 & 963.5414 & 257.2160 & \textbf{256.0309} & & 273.2082 & 257.6278 & 257.7905   & 272.6422 \\
				\addlinespace[0.1cm]
				$\frac{\text{MSE}}{\text{MSE}_{min}}$ & 2.8652  & 2.0023  & 3.7633  & 1.0046  & \textbf{1.0000} & & 1.0671  & 1.0062 & 1.0069 & 1.0648  \\
				\bottomrule
			\end{tabular}%
		}
	\end{subtable}
\end{table}

Results for BM are not reported for the high computational cost involved. Neither MAPE is reported for avoiding divisions by zero. Once again, projection methods were superior to benchmark competitors in real case studies, as shown in Table~\ref{tab:NOX}.

\section{Conclusion}\label{sec:conclusion}

The functional projection methods offer an empirical insight into forecasting functional time series of cyclic nature that repeat curve patterns and dependency patterns between contiguous curves. Both one-step-ahead prediction and dynamic updating are addressed in a unified way, without attending to any statistical model that could explain data generation's mechanism. 

The projection methods were superior to benchmark methods from the functional time series literature for simulated and real data. The Spanish electricity demand curves illustrated how projection methods take into account structural features of the underlying functional time series, such as weekly and yearly periodicity, crucial in the literature of electricity demand forecasting. Particular characteristics of the electricity demand related to the calendar effect and outliers' presence were anticipated by the method for making accurate predictions. Although these peculiarities are known and often used to model and forecast electricity demand, they could be unknown for other data sets. In these cases, the projection methods may be particularly attractive.

According to the data set, the two projection methods were competitive with each other, being on average one better than the other. By construction, they are similar if the $k$-nearest curves surround the most recent functional datum (termed here focal curve), making it deep. However, they differ if this curve is outlying with respect to their nearest curves. The method based on curve envelope projections performs better in the simulation and empirical data analyses in such cases. The superiority is because both punctual and band predictions are computed from curves surrounding the focal from below and above. However, for focal curves inlying with respect to their nearest curves, the method based on $k$-nearest neighbors projections can be even superior, making both methods similar in average, as we observed with the data sets considered in the previous section. This result reveals the common statistical tradeoff between efficiency and robustness that merits a separate study between both methods. Nevertheless, our main goal is to propose a parameter-free forecasting method for functional time series. We remark that the envelope method with $\theta = 1$ presents a competitive and entirely data-driven method. Based on our implementations and executions of the algorithms, the projection methods discussed here lead to substantive improvements in the computational time with respect to the benchmark. \bluetext{Additionally, the method could easily accommodate explanatory variables by-products of a suitable distance between multivariate functions and a multivariate functional depth, already available in the literature \citep{IevaPaganoni2013, pintado2014, claeskens2014, Hubert2017}. Finally, }
we note that the methodology is intuitive and data-driven, making the valuable approach to a broad audience, even without any prior knowledge of functional time series.

\section*{Acknowledgments}
The authors acknowledge insightful comments and suggestions from two reviewers and the Associate Editor. Antonio El\'ias is supported by the Spanish Ministerio de Educaci\'on, Cultura y Deporte under grant FPU15/00625 and the research stay grant EST17/00841. Antonio Elías and Ra\'ul Jim\'enez are partially supported by the Spanish Ministerio de Econom\'ia y Competitividad under grant ECO2015-66593-P. Part of this article was conducted during a stay at Australian National University. Antonio Elías is grateful to Han Lin Shang for his hospitality and insightful and constructive discussions.

\section{Appendix}

\subsection{Functional Autoregressive Processes}\label{sec:appendixFAR}

With the primary aim of validating FPCF as a benchmark method, we repeated the simulation study conducted by \citet{Aue2015}. This study was based on $100$ realizations of eight classes of functional autoregressive (FAR) processes, 800 realizations in total. Each class is characterized by a pair of values $(\kappa_1, \kappa_2)$ and a vector of standard deviations. Two vectors were chosen by the authors, denoted by $(\sigma 1)$ and  $(\sigma 2)$. See Section 6.3 of \citet{Aue2015} for full description of these models. For generating FAR processes data, we used the \texttt{fChange} package \citep{fchangeR}. For each realization, the study considered small $n$ (from 180 to 199) and large $n$ (from 900 to 999). This involves 2000 predictions for small $n$ on each model and 10,000 for large $n$. We chose $q=0.25$. Both for EPF and EPU$_{0.25}$ we considered $m=100$, for small $n$, and  $m=500$, for large $n$. Smaller values of $m$ (20 for small $n$ and 100 for large $n$) were also considered and yield similar results. MSEs produced by the methods mentioned above are shown in Table 1.
\begin{table}[!htbp]
	\setlength{\tabcolsep}{9.5pt}
	\centering
	\begin{tabular}{@{}ccccccccccc@{}}
		\toprule
		&  & \multicolumn{4}{c}{$\sigma_1$} & & \multicolumn{4}{c}{$\sigma_2$}  \\ \cmidrule{3-6} \cmidrule{8-11}
		$\kappa_1$ & $\kappa_2$ & Average & FPCF & EPF & EPU$_{0.25}$ & & Average & FPCF & EPF & EPU$_{0.25}$  \\ \hline 
		0.2 & 0.0 & 1.63 & 1.65 & 1.69 & 0.92 & & 2.30 & 2.33 & 2.42 & 2.02 \\ 
		&  & 1.62 & 1.61 & 1.65 & 0.88 & & 2.30 & 2.30 & 2.36 & 1.91 \\ 
		\addlinespace[0.3cm]
		0.8 & 0.0 & 2.04 & 1.59 & 1.62 & 1.11 & & 2.63 & 2.43 & 2.43 & 2.20 \\ 
		&  & 2.14 & 1.65 & 1.67 & 1.05 & & 2.41 & 2.32 & 2.37 & 2.09 \\ 
		\addlinespace[0.3cm]
		0.4 & 0.4 & 1.92 & 1.74 & 1.84 & 0.70 & & 2.47 & 2.45 & 2.54 & 2.13 \\ 
		&  & 1.83 & 1.63 & 1.74 & 0.97 & & 2.41 & 2.32 & 2.43 & 2.00 \\ 
		\addlinespace[0.3cm]
		0.0 & 0.8 & 2.15 & 1.70 & 2.23 & 1.13 & & 2.64 & 2.45 & 2.75 & 2.27 \\ 
		&  & 2.19 & 1.66 & 2.26 & 1.03 & & 2.57 & 2.36 & 2.67 & 2.11 \\ 
		\bottomrule
	\end{tabular}
	\caption{MSE based on functional average, FPCF, EPF and EPU$_{0.25}$, for the two vector standard deviations, $(\sigma 1)$ and $(\sigma 2)$, chosen by \citet{Aue2015}. The first row of each setting ($\kappa_1,\kappa_2)$ corresponds to small $n$ and the second row to large $n$.}
	\label{tab:1}
\end{table}
From this table and Table 1\footnote{There is a misprint in Table 1 of \citet{Aue2015}: the results for ($\sigma1$) and ($\sigma2$) were switched (personal communication with the authors).} of \citet{Aue2015}, we conclude:
\begin{itemize}
	\item FPCF is competitive with the best method reported by \citet{Aue2015}. 
	\item FPCF has advantages over EPF.
	\item MSE based on DB-U$_{0.25}$ was by far the smallest.
\end{itemize}

We emphasize that FAR processes are not necessarily the kind of FTS that we have discussed at the end of Subsection~\ref{sec:basicNotions}, and consequently, neither EPF nor EPU should work well, although their performance is reasonably good. For the pair $(0.8,0.0)$, EPF was competitive. For these values, the future realization is generated only with the focal curve and innovations. Unlike $ (0.0, 0.8)$, the focal curve is not involved, making EPF a little competitive. For $(0.4, 0.4)$, the difference of MSE based on FPCF and EPF was roughly 0.1, below to 7\% of relative difference. For $(0.2, 0.0)$, the FTS is mainly noise, and the average is the best prediction.

\subsection{Stationary FTS with random shocks} \label{sec:appendixA}

The noise $X$ is generated with a stationary Gaussian process with zero mean and squared exponential autocovariance function
\begin{equation*}
\mbox{Cov}(X(t),X(s)) = \exp(-\frac{|t-s|^2}{2\textit{l}_X^2}).
\end{equation*}
This is the default autocovariance function in Gaussian processes simulation \citep{rasmussen2005}. The parameter $\textit{l}_X$  determines the length of the `wiggles' in the trajectories.

The random function $f$ is a trajectory of a stationary Gaussian process with zero mean and periodic autocovariance function
\begin{equation*}
\mbox{Cov}(f(t),f(s)) =  \exp(-\frac{2\sin^2(\pi |t-s|)}{\textit{l}_f^2}).
\end{equation*}
The parameter $\textit{l}_f$ determines lengthscale in the same way as in the squared exponential autocovariance function. We emphasize that the trajectories of $f$ are periodic functions of period~$1$.

The random shocks process $S_\mu$ is based on two independent processes. First, an occurrence process
\begin{equation*}
\chi_\rho(t) = \sum_{i\geq 1} x_i \mathbbm{1}_{[i-1,i)}(t),
\end{equation*}
$\{x_i, i\geq 0\}$ being a two states Markov chain, with $x_0=0$ and transition probabilities 
$$\mathbb{P}(x_{i+1} = 1|x_i )=\rho(1-x_i), \  \mathbb{P}(x_{i+1} = 0|x_i )=1-\mathbb{P}(x_{i+1} = 1|x_i ).$$ Second, a random periodic function for modeling irregular shock patterns. Specifically, we use $Z(t) = \sigma_g^2(g^2(t)-g^2(0))$, $g$ being an independent copy of $f$. The parameter $\sigma_g$ determines the average distance of $g$ away from zero. The squares are taking for generating a non-zero mean pattern, and $g^2(0)$ is subtracted for beginning to shock from zero. The random shock at time $t$ is then
\begin{equation*}
S_\mu(t) =  \begin{cases}
0, &\mbox{for} \ t <u\\
Z(t-u)\chi_\rho(t-u), &\mbox{for}\  t\geq u\\
\end{cases}
\end{equation*}
$u$ being the minimum time that can elapse before the first shock and \bluetext{$\rho = \mu/(2-\mu)$}. In our simulations, $u$ is uniformly distributed on $[0,1]$. Since the stationary distribution of $\{x_i\}$ is $\mathbb{P}(x_{i} = 1) = \rho/(1+\rho)$ and each shock affects two contiguous periods with probability equals to 1, the proportion of periods affected by shocks, when $t\rightarrow \infty$, is $\mu$.

The parameters considered for examples and simulations are $ \textit{l}_f =1, \textit{l}_X =0.2$, and $\sigma_g^2=10$.

\bibliographystyle{myjmva}
\bibliography{refs}

\end{document}